\documentclass[hidelinks,conference]{IEEEtran}
\usepackage{ifpdf}
\usepackage{amsfonts}
\usepackage{latexsym}
\usepackage{graphicx}
\usepackage{epsfig}
\usepackage{amsbsy}
\usepackage{amsmath}
\usepackage{amssymb}
\usepackage{psfrag}
\usepackage{subfigure}
\usepackage{epstopdf}
\usepackage{microtype}
\usepackage{hyperref}
\usepackage{cite}
\usepackage{color}
\usepackage[T1]{fontenc}
\usepackage[utf8]{inputenc}
\usepackage{tabularx,ragged2e,booktabs,caption}
\newcolumntype{C}[1]{>{\Centering}m{#1}}

\usepackage{multicol}
\usepackage[margin=0.75in]{geometry}

\usepackage{lmodern}

\usepackage{float}

\usepackage{array}

\begin{document}

\title{Revenue Maximization in an Optical Router Node \\
Using Multiple Wavelengths}

\author{\IEEEauthorblockN{Murtuza Ali Abidini\IEEEauthorrefmark{1}\IEEEauthorrefmark{2},
Onno Boxma\IEEEauthorrefmark{1},
Cor Hurkens\IEEEauthorrefmark{1},
Ton Koonen\IEEEauthorrefmark{2}, and
Jacques Resing\IEEEauthorrefmark{1}
}
\IEEEauthorblockA{\IEEEauthorrefmark{1}Department of Mathematics and Computer Science}
\IEEEauthorblockA{\IEEEauthorrefmark{2}Department of Electrical Engineering}
Eindhoven University of Technology\\
P.O. Box 513, 5600MB Eindhoven, The Netherlands\\

\href{mailto:m.a.abidini@tue.nl}{m.a.abidini@tue.nl}, \href{mailto:o.j.boxma@tue.nl}{o.j.boxma@tue.nl}, \href{mailto:c.a.j.hurkens@tue.nl}{c.a.j.hurkens@tue.nl},
\href{mailto:a.m.j.koonen@tue.nl}{a.m.j.koonen@tue.nl}, \href{mailto:j.a.c.resing@tue.nl}{j.a.c.resing@tue.nl}
}

\maketitle

\begin{abstract}
In this paper an optical router node with multiple wavelengths is considered. We introduce revenue
for successful transmission and study the ensuing revenue maximization problem. We present
an efficient and accurate heuristic procedure for solving the NP-hard revenue maximization problem
and investigate the advantage offered by having multiple wavelengths.
\end{abstract}


\IEEEpeerreviewmaketitle

{\let\thefootnote\relax\footnote{{The research is supported by the Gravitation program NETWORKS, funded by the Dutch government.}}}


\section{Introduction}
In the last decades, optical fibers have emerged as the dominant transport medium in communication networks,
because they offer major advantages over copper cables: huge bandwidth, extremely low losses
and an extra dimension, viz., a choice of wavelengths (wavelength division multiplexing).
Multiple wavelengths are to be used in order to enable the packet routing at various planes in the network (each at a specific wavelength); by including wavelength conversion, packets can be transferred between these planes, and thus congestion points can be circumvented. To handle packets at the IP layer would imply lots of packet conversions from optical to electronic, do the IP processing in the electrical domain, and then convert back to optical. The O/E-IP processing-E/O conversions introduce relatively significant time delays, which means extra latency. Such extra latency can seriously reduce the network throughput for interactive high-speed communication between users. All-optical routing in the nodes, as proposed and studied in this paper, is valuable to minimize the latencies.

Optical routing also offers substantial challenges \cite{Maier, Rogiest}.
Photons cannot be stored easily, and hence buffering of optical packets is different from buffering in conventional communication systems.
When photons need to be buffered, they are sent into a local fiber loop, which thus provides a small discrete delay to the photons without displacing or losing them.
Packets can be inserted into and extracted from the fiber loop by means of a cross/bar switch.
If after a loop completion the photon still cannot be transmitted, then it could
again be sent into the fiber loop, or be considered as lost. Such optical nodes are to be used in an all-optical packet-routing network, having multiple hops.

In \cite{Abidini1} we modeled a single-wavelength optical routing node as a queueing system
with a single server (the wavelength) and $N$ stations -- the $N$ ports of the routing node.
We assumed that each successful transmission of a packet ("customer") brings a certain profit.
Our aim in \cite{Abidini1} was to maximize the router performance by maximizing that profit.
As a communication system typically works in frame time, we demanded that the time it takes the server to complete one cycle of the $N$ stations is a given constant $C$.
We then wanted to assign fixed amounts of time $V_1,\dots,V_N$ to the visit periods (also called service windows) of the stations,
such that $\sum_{i=1}^N V_i = C - \sum_{i=1}^N S_i$,
where $S_i$ is the time to switch to station $i \in \{1,2,\dots,N\}$.
We introduced the probability $p_i(V_i)$ that a packet in a retrial loop of station $i$ retries
during visit period $V_i$, and the probability $q_i(V_i)$ that a packet is dropped when it fails to retry during $V_i$.
Under reasonable assumptions on those retry and drop probabilities, the revenue optimization problem in \cite{Abidini1} was shown
to be a separable concave optimization problem -- a well-studied type of optimization problem that allows for an efficient and insightful algorithm
(RANK; cf. \cite{Ibaraki_Katoh}) that
yields the optimal solution.

Our goals in the present paper are (i) to investigate the advantage offered by having {\em multiple wavelengths},
and (ii) to formulate and solve the revenue optimization problem for an optical routing node with multiple wavelengths.
We shall show that the advantage, in terms of revenues, is very significant (in particular, to go from one to two wavelengths).
While solving the revenue optimization problem for multiple wavelengths is an NP-hard problem,
we develop a heuristic that works very well. Our numerical results give insight into the sensitivity of various parameters and modeling assumptions.

%
%

The paper is organized as follows.
Section~\ref{sec:model} presents a detailed description of the optical routing node model.
The revenue maximization problem, that amounts to a
resource allocation problem (assigning stations to wavelengths and assigning visit times to stations),
is discussed in Section~\ref{sec:alloc}. Numerical examples are shown in Section~\ref{sec:num}.
Section~\ref{sec:conclusion} contains conclusions and suggestions for further research.

\section{Optical routing node model}
\label{sec:model}
Consider a $K$-wavelength optical routing node with $N$~ports (stations) to route packets and with retrial loops to store packets. We represent it by a
queueing model with $K$ servers which visit $N$ queues.
We shall assume that there is a fixed assignment of stations to servers (how to do that assignment is part of our optimization problem),
in which each server always serves a fixed set of stations.
\\
{\bf The customers:}
Packets (also called customers in queueing terminology) of type~$j$, $j=1,\cdots,M$,
arrive at station~$i$, $i=1,\cdots,N$, according to independent Poisson processes with rate $\lambda_{ij}$, for all $i,~j$.
We allow several customer types because there can be several types of data at each port.
If at the time of packet arrival the station is being served,
then the packet is instantaneously transmitted; else it enters a retrial loop.
We assume the retrial time to be random, because delay loops of various lengths may be used.
If, at the time of retrial, the station is not in service then the packet again goes into a retrial loop and this process continues.
\\
{\bf The servers:}
The servers go through cycles of fixed length $C$ (the frame time). In each cycle a server visits each of its assigned stations once, for a fixed period of time $V_i$ for station $i$.
A visit to $i$ is preceded by a deterministic switchover (setup) time $S_i$ of the server.
During $V_i$, there may be two types of arrivals: (i) newly arriving packets, and (ii)
packets which were in a retrial loop; we assume the latter retry during $V_i$ with probability $p_i(V_i)$. In view of the huge available bandwidth, we assume the server serves all these packets (new arrivals + retrials) instantaneously, i.e., whenever a station is being served, any packet which arrives at it or retries, is transmitted immediately.
Hence for practical purposes the service times are negligible.
At the end of the visit of station~$i$ each packet which still resides in a retrial loop of $i$ is dropped
with probability $q_i(V_i)$.
Hence the probability
that a packet in a retrial loop of station~$i$ leaves the system, either served during a visit at station~$i$ or dropped after a visit of station~$i$, is $r_i(V_i) := p_i(V_i)+q_i(V_i)-p_i(V_i)q_i(V_i)$.
\\
{\bf Revenue:}
Every served customer generates a profit and every lost customer incurs a loss to the system.
Our goal is to assign stations to servers, and subsequently visit times within a frame time $C$ to stations, such that the revenue of the system is maximized. Assume that:
\begin{itemize}
\item a customer of type $j$ served at station~$i$ gives a profit $\gamma_{ij}$ (depending both on the type of packet and the type of source).
\item a customer of type $j$ dropped at station~$i$ causes a penalty $\theta_{ij}$.
Indeed, the server has an obligation to meet the contract it has with each source. If the server fails to meet this
contract it incurs a penalty: loss of packets/reputation/further contracts.
One could also view $\Theta_i := \sum_j \lambda_{ij} \theta_{ij}$ as contract costs of the service provider per time unit,
and $\Gamma_i := \sum_j \lambda_{ij} (\gamma_{ij} + \theta_{ij})$ as the maximum revenue that can subsequently be earned back by successfully serving customers.
\end{itemize}

For $K=1$ wavelength (cf.\ also \cite{Abidini1} where that case was studied), the mean earnings per cycle are
\[
\sum_j \lambda_{ij} \gamma_{ij} \left[(C-V_i) \frac{p_i(V_i)}{r_i(V_i)} + V_i \right],
\]
and the mean costs per cycle are
\[
\sum_j \lambda_{ij} \theta_{ij} \left[(C-V_i)(1 - \frac{p_i(V_i)}{r_i(V_i)}) \right],
\]
yielding the following net revenue for station~$i$ per cycle:
\begin{equation*}
R_i(V_i) =
M_i(V_i) - C \Theta_i,
\end{equation*}
where for all $i=1,\dots,N$,

\begin{equation}
M_i(V_i) := \Gamma_i \left[(C-V_i) \frac{p_{i}(V_i)}{r_i (V_i) }+ V_i\right].
\label{Revenue}
\end{equation}


In the next section we present an algorithm
to allocate the stations to different wavelengths such that each wavelength has a set of stations to
serve; subsequently the visit periods are chosen such that the revenue for each wavelength is maximized.

\section{Resource allocation}
\label{sec:alloc}
In this section we propose a procedure for solving the revenue maximization problem that was globally described in Section~\ref{sec:model}.
For each wavelength~$k$, we have $C= \sum_{i \in\mathbf{P}_k}(S_i + V_i)$ where $\mathbf{P}_k$ represents the set of all stations served by wavelength~$k$.
Note that if there is only one station being served by a wavelength, then there is no switchover involved. In that case,
$ V_i=C$ where $i$ is the only element of $\mathbf{P}_k$. Further we denote the set of stations which are each served by one complete wavelength as $\mathbf{P}$ and the set
of stations which are not served by any wavelength as $\mathbf{Q}$.

We now define the optimization problem REVENUE which produces maximum revenue via an optimal allocation of stations to wavelengths and visit periods to stations.
\textcolor{blue}{\textbf{REVENUE}}
\begin{eqnarray*}
\text{~~~~~~~max} \sum_{i=1}^{N} M_i(V_i) \\
\text{subject to ~~~~~~~~~~~~~~~~~}\\
\sum_{i=1}^{N} [(S_i+V_i) x_{ik} + V_i y_{ik}] &=& C,~~\forall~k=1,2,\cdots,K ,\\
\sum_{k=1}^{K} [x_{ik}+y_{ik}] &\leq& 1 ,~~\forall~i=1,2,\cdots,N ,\\
\sum_{i=1}^N x_{ik} + N \sum_{i=1}^{N} y_{ik} &\leq& N,~~\forall~k,\\
x_{ik},~y_{ik} \in \{0,1\}&\text{and}& 0 \leq V_i \leq C , ~~~\forall~i,~k.\\
\end{eqnarray*}


Here $M_i(V_i)$ is given in Eq.~\eqref{Revenue}. $x_{ik}=1$ if station~$i$ is served by wavelength~$k$, but station~$i$ is not the only one being
served by it, and is $0$ otherwise, $y_{ik}=1$ if station~$i$ is the only station being served by wavelength~$k$,
and is $0$ otherwise. This is stated in the third condition where if for a wavelength~$k$ some $y_{ik}=1$ then no other
station can be served on it. The second condition states that each station~$i$ can only be served by at most one wavelength.
The first and last conditions are system properties, and they state that the allocation per wavelength
should be equal to its capacity~$C$ and the visit period cannot be negative or more than the total cycle time of one wavelength.
This problem is a non-linear mixed integer programming problem.
Under certain realistic assumptions regarding the system parameters (see also \cite{Abidini1}), we can reduce the objective function of this maximization problem to separable concave terms; however, the occurrence of the integers $x_{ik},~y_{ik}$ prevents us from using the RANK algorithm \cite{Ibaraki_Katoh}
that was used in \cite{Abidini1}. The so-called BALANCE problem, which is NP-complete \cite{Garey},
is a special case of REVENUE. Hence REVENUE is an NP-hard problem;
below we propose a heuristic to solve REVENUE.
We argue that this heuristic should produce results which are close to optimal,
and we provide numerical results in Section~\ref{sec:num} to support that claim.

The idea behind our approach is the following.
In {\it Step~1} we do as if there is only one wavelength, but a frame time of length $KC$.
We use the RANK algorithm to get an optimal choice of the visit periods $\tilde{V}_i$ for such a situation.
That should already give a quite good first estimate of the visit periods.
In {\it Step~2} we use those $\tilde{V}_i$ values to assign stations to wavelengths. This is done such that
each of the $K$ wavelengths gets roughly the same $\sum(S_i+\tilde{V}_i)$ -- which hence should be close to $C$.
Finally, in {\it Step~3}, with those $K$ allocations we use RANK again, but now for $K$ separate single-wavelength problems.
Below we provide the details of these three steps.
\\
\subsubsection*{Step 1}
We first define the following optimization problem.
\\

\noindent
\textcolor{blue}{\textbf{ONE}}
\begin{eqnarray*}
\text{~~~~~~~max}\sum_i M_i(\tilde{V}_i) \\
\text{subject to } \sum_i \tilde{V}_i &=&K C -\sum_i S_i,\\
\text{and~~~~~} 0 \leq ~~~\tilde{V}_i &\leq& C-S_i,~~ \forall i.\\
\end{eqnarray*}


The solution of this optimization problem gives us the values of $\tilde{V}_i$ required by each station to give the maximum revenue, subject to the condition
that the maximum amount of resource available is $K C$. The upper bound on $\tilde{V}_i$ is included because a station cannot be served by more than one wavelength.
Note that $M_i(\tilde{V}_i)$ is the same as given in Eq.~\eqref{Revenue}.

We solve the (separable, concave) optimization problem ONE using RANK, and we thus obtain values of $\tilde{V}_i$. Every station~$i$ which has $S_i +\tilde{V}_i= C$, is allocated
to a single wavelength. These stations belong to the set $\mathbf{P}$ and as described at the start of this section, all stations belonging to this set
have their visit periods equal to the cycle time $C$. Further, all the stations with $\tilde{V}_i=0$ belong to the set $\mathbf{Q}$.
These stations will not be allocated to any wavelength, and as mentioned earlier they will have zero visit period.
By renumbering, we may assume that the stations in $\mathbf{Q}$ are the highest numbered stations, immediately preceded
by the stations in $\mathbf{P}$.
Also assume that the latter $N(\mathbf{P})$ stations are assigned to the $N(\mathbf{P})$ highest numbered wavelengths.
\\

\noindent
We now turn to our procedure for assigning stations to wavelengths (Step 2) and subsequently determining the exact visit periods (Step 3).

\subsubsection*{Step 2}
Take the values of $S_i+\tilde{V}_i $ for the first $N-N(\mathbf{P}+\mathbf{Q})$ stations (i.e., those not in $\mathbf{P}$ or $\mathbf{Q}$).
Sort these values in descending order, say $S_1 + \tilde{V}_1 \geq S_2 + \tilde{V}_2 \geq \cdots \geq
S_{N-N(\mathbf{P}+\mathbf{Q})} + \tilde{V}_{N-N(\mathbf{P}+\mathbf{Q})}$.
Then allocate those stations to the first $K-N(\mathbf{P})$ wavelengths following the so-called {\em Longest Processing Time first} (LPT) rule.
This amounts to first assigning stations $1,\dots,K-N(\mathbf{P})$ to wavelengths $1,\dots,K-N(\mathbf{P})$;
and subsequently assigning each of the remaining stations, one by one in descending order of their values, to that wavelength for which the sum of the already assigned values is the smallest.
This procedure is continued until all stations have been assigned.
\\

{\bf Remark}.
The idea to use LPT comes from multiprocessor scheduling.
Consider a set of $N$ tasks which have to be served on $K$ parallel servers.
The service of a task on a server, once started, cannot be interrupted.
In multiprocessor scheduling the goal often is to minimize the {\em makespan},
i.e., the time until all tasks are completed. This is an NP-hard problem.
The makespan minimization problem can be reformulated in the terminology of bin-packing,
where it amounts to finding the smallest common capacity of the bins, sufficient
to pack all $N$ pieces.
Many heuristics have been developed for solving the bin-packing or makespan minimization problem;
see, e.g., \cite{Coffman}. LPT is a simple and accurate heuristic procedure. It is intuitively clear that assigning tasks in decreasing order of size should work well when $K$ and $N$ are not too small:
because the smallest tasks are assigned last, it is likely that all makespans are close to each other.
See \cite{Ong} for a probabilistic analysis of
various bin-packing heuristics, and \cite{Boxma84} for a probabilistic analysis of LPT list scheduling.
\\
\subsubsection*{Step 3}

Now that we have assigned all stations to a wavelength, we still need to determine the visit periods for those stations that use wavelengths $1,\dots,K-N(\mathbf{P})$,
because the extended visit periods $S_i+\tilde{V}_i$  of the stations  that are assigned to a particular wavelength do not exactly sum up to $C$.
For this we solve optimization problem TWO, for $k=1,\dots,K-N(\mathbf{P})$:
\\

\noindent
\textcolor{blue}{\textbf{TWO}}
\begin{eqnarray*}
\text{~~~~~~~max}\sum_{i \in\mathbf{P}_k} M_i(V_i) \\
\text{subject to } \sum_{i \in\mathbf{P}_k} V_i &=& C -\sum_{i \in\mathbf{P}_k} S_i,\\
\text{and~~~~~~~~~~~~} V_i &\geq& 0 ,~~ \forall i \in\mathbf{P}_k.
\end{eqnarray*}

The solution of this optimization problem gives us the values of $V_i$ required by each station allocated to wavelength~$k$, subject to the maximum
amount of resource available at that wavelength.
We thus obtain new extended visit periods $S_i + V_i$ for stations $1,\dots,N-N(\mathbf{P}+\mathbf{Q})$.
\\

{\bf Remark.} If, in {\it Step 2}, a station~$i^*$ is the only one being assigned to a wavelength, then we do not run TWO for it but take $V_{i^*} =C$.

This concludes the description of the heuristic procedure.
In the next section we shall investigate its accuracy.
Its computational complexity is low.
The optimization problems ONE and TWO are concave separable with linear constraints and can be solved in polynomial time;
and we use ONE once, TWO at most $K$ times. We also use LPT once.
Further, we need to sort the extended visit periods in Step 2 once.

\section{Numerical examples}
\label{sec:num}

In this section we present a few numerical examples to illustrate various properties of our system. For all the examples in this section we assume that
the probability of retrial and drop probability for a station~$i$ are given by $p_i(V_i)=1-e^{-\nu_i V_i}$ and $q_i(V_i)=e^{-\mu_i V_i}$. Further,
the revenue of a station~$i$ is equal to $ M_i(V_i)$ as given in Eq.~\eqref{Revenue}.

\subsection*{Example 1:}
We first consider a toy example with $K=2$ wavelengths and either $N=3$ or $N=4$ stations, for which all possible assignments allocating all stations
to a wavelength are listed.
For each station~$i$, the parameters $\nu_i$ and  $\mu_i$ are equal to $0.5$. The switchover times $S_i = 0.2$ for each station~$i$ and cycle time $C=2$. Finally,
$\Gamma_i=i$, for each station~$i$. The allocation of stations to different wavelengths is shown, along with the corresponding visit period (obtained by using TWO) and the revenue
obtained by the system. Note that an allocation $0$ implies that the station was not allocated to any wavelength.

\begin{table}[!htb]
\centering
\caption{3 station system}
\begin{tabular}{|c|c |c|} \hline
 Allocation& Visit Period & Revenue \\\hline
$\mathbf{[1~~1~~2]}$ & $\mathbf{[0.48  ~~ 1.12  ~~  2.00 ]}$&\textbf{10.11}\\ \hline
$[1~~2~~1]$& $[0.28 ~~ 2.00  ~~  1.32]$ &9.81\\ \hline
$[2~~1~~1]$& $[2.00 ~~ 0.61 ~~   0.99]$&  8.65\\ \hline
 \end{tabular}
 \label{3station}
\end{table}

\begin{table}[!htb]
\centering
\caption{4 station system}
\begin{tabular}{|c|c|c|} \hline
Allocation& Visit Period & Revenue \\\hline
$\mathbf{[0~~1~~1~~2]}$& $\mathbf{[0.00 ~~  0.61~~   0.99 ~~   2.00]}$& \textbf{14.65} \\ \hline
$[1~~2~~2~~1]$ &$[0.14 ~~  0.61  ~~  0.99  ~~  1.46]$&14.25\\ \hline
$[1~~2~~1~~2]$& $[0.28 ~~   0.48 ~~   1.32 ~~  1.12]$&14.03\\ \hline
$[1~~1~~2~~2]$& $[0.48 ~~   1.12 ~~   0.67 ~~   0.93]$ &13.34\\ \hline
$[1~~1~~1~~2]$& $[0.00 ~~   0.61 ~~   0.99  ~~  2.00]$& 14.65\\ \hline
$[1~~1~~2~~1]$& $[ 0.00 ~~   0.48 ~~   2.00  ~~  1.12]$&  14.22\\ \hline
$[1~~2~~1~~1]$& $[ 0.00 ~~   2.00 ~~   0.67 ~~   0.93]$& 13.23\\ \hline
$[2~~1~~1~~1]$& $[2.00 ~~      0.00 ~~   0.67 ~~   0.93]$&11.23\\ \hline
 \end{tabular}
 \label{4station}
\end{table}

In Tables \ref{3station} and \ref{4station} the values given by our procedure described in the
previous section are printed boldface. We observe that in both cases
our procedure gives the best allocation.

\subsection*{Example 2:}
In this example we compare the results obtained using our procedure
with the results obtained by randomly allocating wavelengths to different stations
and then optimizing the visit periods at each wavelength. We show numerical results for five different cases for a system
with $N=16$~stations, $K=4$~wavelengths and frame time $C=8$.
In each of the first four cases, we vary one parameter while keeping all the other constant and
in the last case we use random system parameters;
the $\Gamma_i$ are uniformly distributed on $(0,8)$; the $\nu_i$ and $\mu_i$ on  $(0,1)$,
and the $S_i$ on $(0,0.4)$.

We take 10000 independent allocations of wavelengths in two different ways, (i) and (ii). In (i) we allocate stations
in such a way that each wavelength gets at most $4$ stations, whereas in (ii) there is
no restriction on the number of stations allocated to a wavelength.
In both cases we subsequently use TWO.
For both (i) and (ii) we show the maximum, the average and the minimum obtained revenue among the $10000$ cases and
the percentage of allocations which generated a revenue above the value generated using our algorithm.

\begin{table}[!htb]
 \centering
 \caption{Varying $\Gamma_i$}
 \begin{tabular}{|c|c c c|c|} \hline
~& Maximum & Average &Minimum &Percent \\ \hline
 (i)& 475.72&468.89&454.24&1.46 \\ \hline
  (ii)&475.50& 441.36&300.33&0.24\\ \hline
  Algorithm&~&474.51&~&~\\ \hline
  \end{tabular}
  \caption*{$\Gamma_i =0.5*i$, $\nu_i=0.5$, $\mu_i=0.5$ and $S=0.2$.}
\label{varygamma2}
  \end{table}


\begin{table}[!htb]
 \centering
 \caption{Varying $\nu_i$}
 \begin{tabular}{|c|c c c |c|} \hline
~& Maximum & Average&Minimum &Percent \\ \hline
  (i)& 387.29&384.58&381.94&9.89 \\ \hline
  (ii)&387.14& 358.36&224.93&0.87\\ \hline
  Algorithm&~&385.65&~&~\\ \hline
  \end{tabular}
  \caption*{$\Gamma_i =4$, $\nu_i=0.05*i$, $\mu_i=0.5$ and $S=0.2$.}
\label{varynu2}
  \end{table}


\begin{table}[!htb]
 \centering
 \caption{Varying $\mu_i$}
 \begin{tabular}{|c|c c c|c|} \hline
~& Maximum & Average& Minimum &Percent \\ \hline
  (i)& 413.19&413.15&412.98&0.00 \\ \hline
  (ii)&413.19&377.54&  231.52&0.00\\ \hline
  Algorithm&~&413.19&~&~\\ \hline
  \end{tabular}
  \caption*{$\Gamma_i =4$, $\nu_i=0.5$, $\mu_i=0.05*i$ and $S=0.2$.}
\label{varymu2}
  \end{table}


\begin{table}[!htb]
 \centering
 \caption{Varying $S_i$}
 \begin{tabular}{|c|c c c|c|} \hline
~& Maximum & Average&Minimum &Percent \\ \hline
  (i)& 398.81&  398.06&395.60&0.05 \\ \hline
  (ii)&398.79&351.53&181.94&0.00\\ \hline
  Algorithm&~&398.81&~\\ \hline
  \end{tabular}
  \caption*{$\Gamma_i =4$, $\nu_i=0.5$, $\mu_i=0.5$ and $S=0.05*i$.}
\label{varys2}
  \end{table}


\begin{table}[!htb]
 \centering
 \caption{Completely Random}
 \begin{tabular}{|c|c c c|c|} \hline
~& Maximum &Average& Minimum &Percent \\ \hline
  (i)&   360.85&  355.23 &  338.07&4.56 \\ \hline
  (ii)&360.83&338.14& 231.45&0.62\\ \hline
  Algorithm&~&359.93&~&~\\ \hline
  \end{tabular}
  \caption*{$\Gamma_i \sim U(0,8)$, $\nu_i \sim U(0,1)$, $\mu_i \sim U(0,1)$, and $S_i \sim U(0,0.4)$}
\label{varyrandom2}
  \end{table}

Tables \ref{varygamma2}-\ref{varyrandom2} suggest that a random assignment of stations to wavelengths, but still using TWO to subsequently
choose $V_i$, is much worse than the assignment of our algorithm.
However, the symmetric assignment, in which each of the four wavelengths serves (at most) four out of the $16$
stations, and for which the visit times are calculated using TWO, yields results that are typically quite close to the values
obtained using our algorithm (and in a few cases even better).

\subsection*{Example 3:}
In this example we study which effect increasing the number $K$ of wavelengths has on the revenue of the system.
We take the allocation obtained using the procedure of Section~\ref{sec:alloc}. For each $K$ we take $N=16$~stations,
$S_i=\mu_i=\nu_i=0.05*i$, $\Gamma_i = 0.5*i$ and $C=8$.

\begin{table}[!htb]

\centering
 \caption{Varying the number of wavelengths}
 \begin{tabular}{|c|c|c|}\hline
 $K$&Revenue& \# of stations served \\\hline
 1& 170.54 & 3 \\\hline
 2& 322.62& 8 \\\hline
 3& 400.97& 11 \\\hline
 4&452.88& 13 \\\hline
 5&480.40& 14 \\\hline
 6&499.60& 14 \\\hline
 7&517.23& 15 \\\hline
 8&525.21& 15 \\\hline
 16&544.00&16\\\hline
  \end{tabular}

\end{table}

We observe that increasing the number of wavelengths increases the revenue obtained and also the number of stations served.
However, the marginal increment decreases with an addition of each wavelength. In this example the change from $K=1$ to $K=2$ almost doubles the revenue and more than
doubles the number of stations served, whereas the change from $K=7$ to $K=8$ increases the revenue by less than two percent (and the number of stations served
does not change). In the case of $K=16$, the revenue equals $C*\sum_{i=1}^{16} \Gamma_i=544$. The system
operator can choose an optimal number of wavelengths  so as to maximize its utility.
This observation may be of interest in networks where traffic is highly variable and the cost of running extra resources is high.

\subsection*{Example 4:}

In this example we consider a system with $N=16$~stations, $K=4$~wavelengths, frame time $C=8$ and switchover period from each station $S_i=0.2$, for all $i=1,\ldots,N$.
We show three different cases,
each of which has one of $\Gamma_i$, $\nu_i$, and $\mu_i$ different for all stations, the other two
parameters being equal for all stations. In these numerical experiments we study how the procedure described
in Section~\ref{sec:alloc} allocates resources depending on each factor, and develop insight into the influence of
these factors on the system performance.
In Table~\ref{varyall1}, we mention the wavelength to which each station is
assigned, the visit period each station receives and the revenue each station gives, for the three cases.

\begin{table*}[htb]
\centering
 \caption{Visit period allocation and corresponding revenue obtained}
 \begin{tabular}{|c|}\hline
Station\\\hline
1\\\hline
2\\\hline
3\\\hline
4\\\hline
5\\\hline
6\\\hline
7\\\hline
8\\\hline
9\\\hline
10\\\hline
11\\\hline
12\\\hline
13\\\hline
14\\\hline
15\\\hline
16\\\hline
Total\\\hline
 \end{tabular}
 ~~~~~~~~~
 \begin{tabular}{|c c c|}\hline
 Allocation&Visit&Revenue\\\hline
 0&0.00&0.00\\\hline
 0&0.00&0.00\\\hline
 3&0.93&6.54\\\hline
 4&1.22&10.68\\\hline
 4&1.45&14.89\\\hline
 3&1.67&19.27\\\hline
 2&2.16&24.96\\\hline
 1&2.25&28.90\\\hline
 1&2.34&32.89\\\hline
 2&2.46&37.00\\\hline
 3&2.20&39.45\\\hline
 4&2.23&43.23\\\hline
 4&2.30&47.24\\\hline
 3&2.40&51.49\\\hline
 2&2.78&57.03\\\hline
 1&2.81&60.94\\\hline
 ~&29.20&474.51\\\hline
 \end{tabular}
 ~~~~~~~~~
 \begin{tabular}{|c c c|}\hline
 Allocation&Visit&Revenue\\\hline
  0&0.00&0.00\\\hline
  1&3.35&26.05\\\hline
  2&2.33&22.33\\\hline
  3&2.18&23.09\\\hline
  4&2.07&23.83\\\hline
  4&1.97&24.37\\\hline
  3&1.88&24.80\\\hline
  2&1.83&25.30\\\hline
  1&2.16&28.02\\\hline
  4&1.69&25.85\\\hline
  3&1.64&26.09\\\hline
  2&1.60&26.42\\\hline
  1&1.89&28.59\\\hline
  3&1.50&26.76\\\hline
  4&1.47&26.96\\\hline
  2&1.44&27.19\\\hline
  ~&29.00&385.65\\\hline
 \end{tabular}
 ~~~~~~~~~
 \begin{tabular}{|c c c|}\hline
  Allocation&Visit&Revenue\\\hline
  3&1.85&22.76\\\hline
  4&1.86&23.36\\\hline
  2&1.87&23.94\\\hline
  1&1.87&24.48\\\hline
  3&1.86&24.90\\\hline
  4&1.85&25.29\\\hline
  2&1.84&25.66\\\hline
  1&1.83&26.01\\\hline
  1&1.82&26.32\\\hline
  3&1.80&26.56\\\hline
  2&1.78&26.81\\\hline
  4&1.76&27.03\\\hline
  4&1.73&27.23\\\hline
  2&1.71&27.43\\\hline
  3&1.69&27.62\\\hline
  1&1.68&27.79\\\hline
  ~&28.80&413.19\\\hline
 \end{tabular}

\vspace{0.2cm}
~~~~~~~~~~~~~~~
 \begin{tabular}{c}
  (a)~$\Gamma_i =0.5*i$, $\nu_i=0.5$ and $\mu_i=0.5$.
 \end{tabular}
  \begin{tabular}{c}
  (b)~$\Gamma_i =4$, $\nu_i=0.05*i$ and $\mu_i=0.5$.
 \end{tabular}
 \begin{tabular}{c}
 (c)~$\Gamma_i =4$, $\nu_i=0.5$ and $\mu_i=0.05*i$.
 \end{tabular}
\label{varyall1}
 \end{table*}

From Table \ref{varyall1}(a) we see that in general $\Gamma_i > \Gamma_j$ does not imply $V_i > V_j$, but when $i$ and $j$ are allocated
to the same wavelength this implication appears to be true.
Also, if the value of $\Gamma_i$ is very low, then --  even though our procedure allocates that station to a wavelength --
it may not receive any service (equivalent to not being allocated).

In Table \ref{varyall1}(b) we see that in general, within a wavelength, stations with lower $\nu_i$ receive higher $V_i$.
This happens because the system tries to allocate longer visit periods to stations with low retrial rates so as to maximize the number
of customers it can serve. However, if $\nu_i$ is very low (see station 1), then the system, subject to limited resources, might not allocate any resource to that station.

From Table \ref{varyall1}(c) one can generally observe that the stations with higher drop probability,
i.e., lower $\mu_i$, receive higher visit periods to have fewer losses. Also, like in the previous
case the difference in revenue generated from each station is not big.

 Three final observations:
1.~The spread in visit periods is small in \ref{varyall1}(c) compared to those in \ref{varyall1}(a) and \ref{varyall1}(b).
This suggests that the factor $\mu_i$ is less important than the factors $\nu_i$ and $\Gamma_i$
in the solution of this problem.
2.~Our procedure often results in a more or less even spread of revenues among stations if $\Gamma_i$ are equal.
This suggests that the procedure makes the system reasonably fair, i.e., tries to provide the best service to each station.
3.~Even though the revenues obtained from stations with different retrial rates and drop probabilities are similar, the resources required by these stations are different. For a lower
retrial rate and/or higher drop probability, a longer visit period is required to give similar revenue. This is a techno-economic trade-off to consider while designing the router.

\section{Conclusions and suggestions for further research}
\label{sec:conclusion}

To understand the behaviour and study the performance of future optical networks, we have considered a revenue optimization problem for a multiple wavelength optical routing node.
This is a mixed integer non-linear programming problem and hence extremely time-consuming to solve even for a small number of wavelengths. Since one would like to solve this revenue
optimization problem quite frequently, we have developed an efficient and near-optimal heuristic procedure for
(i)~assigning stations to wavelengths and subsequently (ii)~assigning visit times to stations within a fixed frame time.

Several topics for further research suggest themselves. Firstly, one might consider variants of the proposed heuristic procedure.
For example,
a consequence of the use of LPT is that, for each wavelength, one has a sum of assigned $S_i+V_i$
that is not exactly equal to $C$.
We subsequently used TWO
to make final choices for the visit periods $V_i$.
Instead, one could simply scale all $V_i$, that belong to one and the same wavelength, by the same factor $\alpha$ such that
$\sum(S_i+\alpha V_i) = C$.
Secondly, it could be interesting to
relax two modeling assumptions, viz.,
to take the finiteness of the fiber loops into account
more explicitly, and to remove the assumption of negligible service times.
Thirdly, one could consider completely different ways of assigning stations to wavelengths, allowing for example
that the same station uses more than one wavelength. Finally, it would be worthwhile to study the trade-off between using more wavelengths
and investing in a higher number of fiber loop buffers - which can be translated into a lower drop probability, in terms of node throughput and economics.

\end{document}